\documentclass[prd,aps,floats,twocolumn]{revtex4}
\begin{document}
\newcommand{\beq}{\begin{equation}}
\newcommand{\eeq}{\end{equation}}
%
% % %jornais
%
\newcommand{\Prd}{Phys.  Rev. D$\;$}
\newcommand{\Prl}{Phys.  Rev.  Lett.}
\newcommand{\Plb}{Phys.  Lett.  B}
\newcommand{\Cqg}{Class.  Quantum Grav.}
\newcommand{\Np}{Nuc.  Phys.}
\newcommand{\Grg}{Gen.  Rel.  \& Grav.}
\newcommand{\Fp}{Fortschr.  Phys.}
\newcommand{\Sch}{Schwarszchild$\:$}
\renewcommand{\baselinestretch}{1.2}

\title{Analysis of static and spherically-symmetric solutions in NDL theory of gravitation}

\author{M. Novello, S. E.  Perez Bergliaffa}
\address{Centro Brasileiro de
Pesquisas Fisicas, Rua Xavier Sigaud, 150, CEP 22290-180, Rio de
Janeiro, Brazil}
\author{K. E. Hibberd}
\address{Departamento de F\'{\i}sica Te\'{o}rica, Universidad de Zaragoza, 50009 Zaragoza, Spain}
%\date{\today}
\vspace{.5cm}

\begin{abstract}
We investigate a static solution with spherical symmetry
of a recently proposed field theory of gravitation. In this so-called NDL theory,
matter interacts with gravity in accordance with the Weak Equivalence Principle, while
gravitons have a nonlinear self-interaction.
It is shown that
the predictions of NDL agree with those of
General Relativity in the three
classic tests. However, there are potential differences
in the strong-field limit, which we illustrate by proving that this theory
does not allow the existence of static and spherically
symmetric black holes.
\end{abstract}
\date{4 June, 2002}

\vskip2pc
\maketitle

\section{Introduction}
The Equivalence Principle played a fundamental role in the development of
General Relativity (GR) and is at the heart of the idea that spacetime is curved \cite{will}.
However, a distinction  must be made
between the Weak Equivalence Principle (WEP) and the
Strong Equivalence Principle (SEP). In a nutshell, WEP states
that all matter fields (except the gravitational field) interact with
gravity in the same way (the so-called ``universal coupling'').
SEP states, among other things, that all fields (including gravity) interact
with gravity in the same way. Let us remark that while the other aspects of SEP
(i.e. the absence of preferred-frame and preferred-location effects) have been
tested in several instances \cite{pap1}, there is no conclusive evidence about the way that
gravity couples to gravity. This is precisely the area explored by a
recently presented field theory of gravitation, called NDL by the authors of \cite{pap1}.
This theory is a field-theoretical description
of gravity (in the spirit of Feynman \cite{fey} and Deser \cite{deser}) in which
SEP is violated from the beginning:
gravity does not couple to itself in the same way it couples to other fields.

Several features of this theory have been studied in \cite{vel1,vel2,vel3}.
To date, the most important prediction of NDL theory is that gravitational
waves do not propagate at the same speed as
electromagnetic waves \cite{vel1}, a phenomenon related to the
self-coupling of gravity in this
theory. In fact, it can be shown that gravitons in NDL theory  follow
an effective metric, different to the background metric \cite{pap1}, that
depends on the abovementioned self-coupling. Indeed,
this is a feature of any nonlinear theory: the influence of the
effective geometry on nonlinear photons has been extensively analyzed in \cite{nlem1,emwh}.

To go further with the analysis of the predictions of NDL theory,
in this article we shall study some aspects
of the static and spherically symmetric solution obtained in \cite{pap1}. We begin
by giving a short summary of NDL theory in
Sect.\ref{summary}. The spherically symmetric and static solution
is presented in Sect.\ref{ssss}.
We study in Sect.\ref{co}
some aspects of gravitation
around compact objects.
Specifically,
we investigate the effective potential felt by particles
moving in this solution.
Armed with the effective potential, we compare
the predictions of NDL with those of GR in the three classic tests.
We also
investigate
if the spherically symmetric and static solution can describe a black hole
analog to that of \Sch\/ in GR, and analyze the effective metric felt by gravitons.
We close in Sect.\ref{disc}
with some comments
regarding the nature of the singularities appearing in the solution and prospects for
future work.

{\bf get rid of eqn numbers}

\section{Summary of NDL theory}
\label{summary}
A detailed presentation of NDL theory was given in \cite{pap1}. Here we shall list
the most important features of the theory, following Ref.\cite{vel1}:

\begin{itemize}

\item The gravitational interaction is represented by a symmetric tensor $\varphi_{\mu\nu}$ that
obeys a nonlinear equation of motion.

\item Matter (but not gravity)
couples to gravity through the metric $g_{\mu\nu} = \gamma_{\mu\nu}+\varphi_{\mu\nu}$,
where $\gamma_{\mu\nu}$ is the flat background metric.

\item The self-interaction of gravity, given by a nonlinear Lagrangian, breaks the
interpretation of gravity as a universal geometric phenomenon. That is, all
particles (except gravitons)
move following geodesic of $g_{\mu\nu}$. Gravitons instead move on geodesics of an effective
metric, the expression of which we shall give in Sect.(\ref{effmetricg}).

\end{itemize}
It is convenient to define the tensor $F_{\alpha\beta\mu}$ (called the gravitational field), in
terms of $\varphi_{\mu\nu}$ as follows \footnote{We are using the index
notation $[x,y] = xy-yx$, and $(xy) = xy+yx$.}:
$$
F_{\alpha\beta\mu} = \frac 1 2 \left( \varphi_{\mu [\alpha ;\beta ]} +
F_{[\alpha}\gamma_{\beta]\mu}\right),
$$
where the covariant derivative is constructed with the background metric $\gamma_{\mu\nu}$.
Indices are raised and lowered with that metric also, and
$$
F_\alpha \equiv F_{\alpha\mu\nu}\gamma^{\mu\nu}.
$$
To construct a nonlinear theory of the gravitational field $F_{\mu\nu\alpha}$
with the correct weak field limit,
we assume that the interaction of gravity with itself is described by a functional of
$A-B$, where the scalars $A$ and $B$ are given by:
$$
A= F_{\alpha\beta\mu}F^{\alpha\beta\mu},
\;\;\;\;\;\;\;\;\;\;B = F_\alpha F^\alpha .
$$
From the action
$$
S= \int d^4x\;\sqrt{-\gamma}\; {\cal L}(A-B) ,
$$
(where $\gamma$ is the determinant of the background metric)
and using that fact that ${\cal L}_A = -{\cal L}_B$
(where ${\cal L}_X$ is the derivative of the Lagrangian with respect to $X$),
we get the equations of motion
\beq
\left(\sqrt{-\gamma} {\cal L}_A F^\lambda_{\;(\mu\nu)}\right)_{;\lambda} = 0.
\label{eom}
\eeq
In the next section we shall study a particular solution of these equations
of motions for a given choice of the Lagrangian.

\section{The static and spherically symmetric solution}
\label{ssss}

In what follows, we will restrict our study
to a special Lagrangian studied previously in \cite{pap1},
and inspired by
the Born-Infeld Lagrangian \cite{bi}:
\beq
{\cal L}(A-B) = \frac{b^2}{\kappa}\left( \sqrt{1-\frac{A-B}{b^2}}-1\right)
,
\label{bill}
\eeq
where $\kappa$
is Einstein's constant. The parameter $b$, with dimensions of length$^{-1}$,
is undetermined at this point. Notice that its value should be large enough
for the series expansion of Eqn.(\ref{bill}) to be in agreement with the
weak-field limit.

We are interested in static and spherically symmetric solutions in a Minkowskian background.
Consequently, the only nonzero
components of the field $\varphi_{\mu\nu}$ are
$$
\varphi_{00}\equiv \mu (r), \;\;\;\;\;\;\;\;\varphi_{11} \equiv -\nu (r).
$$
and the corresponding nonzero components of the tensor $F_{\alpha\beta\mu}$ are
given by
$$
F_{100} = -\frac{\nu}{r},
$$
$$
F_{122}= \frac 1 2 (\nu r - \mu ' r^2 ),
\label{f}
$$
$$
F_{133} = F_{122}\sin^2\theta.
$$
The only nonzero trace component, $F_1$, is given by
$$
F_1 = \mu ' -\frac 2 r \nu.
$$
From the equations of motion (\ref{eom}) and the expression of the Lagrangian
we get only two nontrivial equations:
$$
2\nu^3 + b^2 r^3\nu ' + b^2 r^2 \nu = 0,
$$
$$
\mu ' r - \nu = 0
%\label{eom2}
$$
The first equation can be easily integrated,
and the result is
\beq
\nu (r) =
\epsilon\; \frac C r \left[ 1 - \left( \frac{r_0}{r}\right) ^4\right] ^{-1/2} ,
\label{nu}
\eeq
where we have defined
\beq
r_0 ^2 = \frac {C}{|b|},\;\;\;\;\;\;\;\;
\epsilon = \pm 1,
\eeq
and $C$ is an integration constant. From the second equation,
$$
\mu (r) = \int \frac{\nu (r)}{r}\; dr + {\rm const.}
$$
Noting that the function $\nu (r)$ is defined only for $r\geq r_0$,
we can write using Eqn.(\ref{nu}),
$$
\mu (r)= \epsilon \;C \int_{r_0}^r \frac{dr}{\sqrt{r^4 - r_0^4}} + {\rm const.}
$$
This integral can be written in terms of an elliptic integral of the first kind
\cite{rtable} using the identity
$$
\int _\beta^u \frac{dx}{\sqrt{(x^2+\alpha^2)(x^2-\beta^2)}} = \frac{1}{\sqrt{\alpha^2 +
\beta^2}}
\;F(X,Y),
$$
valid for $u>\beta>0$, and
$$
X = \arccos\left(\frac \beta u\right) ,\;\;\;\;\;\;\;\;\;\;\;
Y = \frac{\alpha}{\sqrt{\alpha^2+\beta^2}} .
$$
Consequently, $\mu(r)$ is given by
\beq
\mu  (r) = \frac{\epsilon \;C}{\sqrt 2\; r_0} F(\arccos (r_0/r), 1/\sqrt 2) + {\rm const.}
\eeq
In order to determine the value of the constant, we impose that spacetime be
Minkowskian for large $r$, which means that $\mu (r) \rightarrow 0$ in that limit. Then,
\beq
\mu (r) =\epsilon \sqrt{\frac{|b|C}{2}} \left[ F(\arccos (r_0/r), 1/\sqrt 2)
- F(\pi/2, 1/\sqrt 2 ) \right] .
\label{mu}
\eeq
The value of the constant $C$  and the
sign of $\epsilon$ have not been determined up to now: they
are dictated by the
weak field limit of the solution, which must coincide with that of Schwarszchild.
The radial component of the Schwarszchild metric (we are setting $G=c=1$) is
$$
g_{rr}(r) = -1 - \frac{2M}{r}
.
$$
In our case, $g_{rr} (r) = -1-\nu(r)$, and in the weak field limit we get
\beq
g_{rr}(r) = -1 - \epsilon \frac{C}{r}
,
\label{dev}
\eeq
and therefore conclude that $\epsilon = +1$ and $C = 2M $.

Thus, the general expression for the static,
spherically symmetric and asymptotically flat
spacetime in
NDL theory previously derived in \cite{pap1}, is given by the metric
\beq
ds^2 =
[1+\mu (r) ]\; dt^2 - [1 + \nu (r) ]\;dr^2 - r^2 d\Omega^2 ,
\label{metric}
\eeq
with $\mu (r)$ and $\nu (r)$ given by Eqns. (\ref{nu}) and (\ref{mu}) respectively.
In the following sections we
shall study
some properties of this
solution.

\section{Gravitational physics around compact objects in NDL theory}
\label{co}
Here the predictions of NDL theory are compared with those of GR.
Using the static and spherically symmetric solution, we
will consider first the three
classical tests of GR in the framework of NDL theory.
Then
we shall see if NDL theory can
describe black hole configurations analog to that of Schwarszchild.
The following series
expansions will be necessary in the ensuing sections.
\beq
g_{tt} (r) =
1- \frac Cr \left[ 1 + \frac {1}{10}\left(\frac{r_0}{r}\right)^4 + \frac{1}{24}
\left(\frac{r_0}{r}\right)^8 + ...\right],
\label{gtt}
\eeq
with $r_0^2=2M/|b|$. The inverse of $g_{tt}$ and $g_{rr}$ are given by
\beq
g_{tt}^{-1}(r) = 1 + \frac C r + \frac {C^2} {r^2} + \frac {C^3} {r^3} + O(r^{-4}),
\label{d1}
\eeq
\beq
g_{rr}^{-1}(r) = -1 + \frac C r - \frac {C^2} {r^2} + \frac {C^3} {r^3} + O(r^{-4}).
\label{d2}
\eeq

\subsection{Motion of particles}

We now analyze the geodesics of photons and particles with nonzero mass
in the geometry described by Eq.($\ref{metric}$).
The motion can be studied using the effective potential \cite{emwh,wald}.
For a
static
and spherically symmetric geometry there are two constants of
motion along a
geodesic, which we designate by $E$ and $L$:
\beq
g_{\phi\phi} \dot\phi = L ,\;\;\;\;\;\;\;\;\;\;\;\;\;\; g_{tt}\dot t = E
,
\label{cons}
\eeq
where the dot indicates a derivative with respect to the proper time (or an
affine parameter in the case of null geodesics), and $\theta = \pi/2$.
From the equation
for the interval it follows that
\beq
k = g_{tt} \dot t^2 + g_{rr} \dot r^2 +g_{\phi\phi} \dot
\phi^2 ,
\label{int}
\eeq
where $k$ is 1 (0) for timelike (null) geodesics.  Rearranging
Eqns.(\ref{cons}) and (\ref{int}) we obtain
\beq
\dot r^2
+V(r) = E^2
,
\label{rdot}
\eeq
where the effective potential $V(r)$ is given by
$$
V(r) = -\frac{1}{g_{rr}(r)}
\left[ \frac{L^2}{r^2} + k \right] + E^2 \left(\frac{1}{g_{tt}(r)g_{rr}(r)} + 1
\right)
,
%\label{effpot}
$$
with $g_{rr}(r)$ and $g_{tt}(r)$ defined in Eqn.(\ref{metric}). $L$ and $E$ are the angular
momentum and energy per unit mass respectively.
This expression for $V(r)$
reduces to the Schwarszchild case of GR,
\beq
V_{\rm Sch} (r)= k \left( 1 - \frac{C}{r}\right) + \frac{L^2}{r^2}
- \frac{CL^2}{r^3},
\label{grpot}
\eeq
upon substitution of the \Sch metric.

\subsubsection{{\bf The perihelion shift of Mercury}}

One of the triumphs of GR was its ability to correctly predict the perihelion of
Mercury. We shall show here that
NDL concurs with this prediction.
The values
$$
C_\odot \approx  3{\rm km}, \;\;\;r_+ \approx 7.0\times 10^7 {\rm km}, \;\;\;\;\;
r_-\approx 5.0\times 10^7 {\rm km},
$$
will be used below, where $r_+$ ($r_-$) is the aphelion (perihelion) of Mercury.
With these values, we see that
$$
\left(\frac{r_0}{r_\pm}\right)^4\approx \frac{1}{|b|^2}\;10^{-28}.
$$
This equation and  the discussion following Eqn.(\ref{bill}) ensure that we can
keep only the first term in the expansion in Eqn.(\ref{gtt}). Notice also that
$$
\left(\frac{C_\odot}{r_\pm}\right)^4 \approx 10^{-29},
$$
and consequently
from the expansions in Eqns.(\ref{d1}) and (\ref{d2})
the fourth and higher orders terms
can be neglected.
The effective potential in the case under consideration can thus be approximated by
\beq
V(r) = k\left(1 - \frac Cr\right) +\frac{L^2}{r^2} - \frac{CL^2}{r^3}\left(
1+k\frac{C^2}{L^2}\right) + \frac{C^2}{r^2}( k - E^2 ).
\label{appeffpot}
\eeq
We see that this approximate expression contains new terms not present
in the exact expression given by Eqn.(\ref{grpot}), even for the massless case.
These new terms signal
potential differences
between NDL theory and GR in the weak field regime. Let us try to estimate the magnitude
of these terms for the case of Mercury, in which $C^2/L^2\approx 10^{-7}$, so
the term involving this factor
can be safely neglected. We now need an estimate for $E^2$
which may be obtained using
Eqn.(\ref{rdot}) evaluated at $r_+$ and $r_-$ (the values of the radius
where $\dot r = 0$).
The result can be written as
a system of two equations in the unknowns $E^2$ and $L^2$. In particular,
$$
E^2 = \frac{g_{tt}(r_+)g_{tt}(r_-)\; (r_+^2 - r_-^2)}{r_+^2 g_{tt}(r_-) -
r_-^2 g_{tt}(r_+)}.
$$
It follows that
$E^2 - 1 \approx 10^{-7}$ and so the last term in Eqn.(\ref{appeffpot})
is negligible for $k=1$. Consequently the effective potential
for NDL theory coincides with that of
Schwarszchild. Therefore,
NDL agrees with GR in the prediction for the perihelion of Mercury.

\subsubsection{\bf The deflection of light by the Sun}
To derive the equation that governs the deflection of light rays by the Sun,
we use the fact that
\beq
\frac{C_\odot}{ R_\odot} \approx 4.23\times 10^{-6} .
\label{app}
\eeq
Combining this with the discussion following Eqn.(\ref{bill}),
it follows that the effective potential for the deflection of light
is given by Eqn.(\ref{appeffpot}) with $k=0$. That is,
$$
V(r) = \frac{L^2}{r^2}\left( 1 - \frac{C}{r}\right)
- \frac{C^2}{r^2} E^2.
\label{appeffpot2}
$$
Consequently, the equation of motion takes the form
$$
\dot r^2 + \frac{L^2}{r^2}\left( 1 - \frac{C}{r}\right) = \left[ 1 +\left(\frac{C}{r}\right)^2
\right] E.
$$
From Eqn.(\ref{app}) we have that $(C/r)^2<10^{-13}$, which is negligible.
Thus it is clear that the equation
of motion for photons in NDL coincides with that of Schwarszchild.

\subsubsection{\bf Time delay of light}

Let us sketch the derivation of this effect in NDL theory, adopting the usual
procedure for GR (after \cite{dinverno}). Consider the path of a light ray with
$\theta = \pi/2$ in the metric given by Eqn.(\ref{metric}):
$$
g_{tt}dt^2 + g_{rr}dr^2 - r^2 d\phi^2 = 0.
$$
This equation can be rewritten as
$$
dt^2 = dr^2\left[ r^2 \left( \frac{d\phi}{dr}\right)^2 g_{tt}^{-1} - g_{rr} g_{tt}^{-1}
\right].
$$
Using the expansions given in Eqns.(\ref{dev}) and (\ref{d1})
we get
$$
dt^2 = dr^2\left[ r^2 \left( \frac{d\phi}{dr}\right)^2 \left( 1 + \frac C r \right) +
1 + \frac{2C}{r} \right],
$$
which coincides with the equation for the GR case (see \cite{dinverno}, pp. 204).
Consequently, also in this case NDL theory prediction agrees with that of GR.

\subsection{Black holes in NDL}

We now investigate the existence of horizons in the metric given by Eqn.(\ref{metric}).
The position of the putative horizons is given by the condition $g_{tt} (r_h) = 0$.
This
implies
$$
F(\arccos (r_0/r_h),
1/\sqrt 2) - F(\pi/2, 1/\sqrt 2 ) = - \sqrt 2\; \frac{r_0}{C}.
%\label{hor}
$$
By standard manipulations (see for instance \cite{rtable}) we obtain
%\beq
%u \equiv F(\phi, k) = \int _0^\phi
%\frac{d\alpha}{\sqrt{1-k^2\sin ^2\alpha}} \equiv u(\phi) ,
%\eeq
%and we can define the
%inverse function
%\beq \phi = {\rm am}\;u
%\label{am}
%\eeq
%If we start with the other integral instead, {\em i.e.}
%\beq
%u \equiv F(\phi, k) = \int _0^x
%\frac{dt}{\sqrt{(1-t^2)(1-k ^2 t^2)}} \equiv u(\phi)
%\eeq
%with $x = \sin\phi$, the inverse function is {\bf different}:
%\beq
%\sin \phi = {\rm sn}\; u
%\label{inv}
%\eeq
%or
%\beq
%\phi = \arcsin {\rm sn}\; u
%\eeq
%So there are two different integral representations for the function $F(\phi , k)$,
%and from each of them we can define an inverse wrt the upper limit of
%integration. Of course, to be consistent we must have
%\beq
%{\rm am}\; u = \arcsin {\rm sn}\; u
%\eeq
%and this equation is in the  russian table, p. 910.

%From Eqn.(\ref{hor}) we get
%\beq
%F(\arccos (r_0/r_h), 1/\sqrt 2) =  F(\pi /2,
%1/\sqrt 2 ) - \epsilon \sqrt{\frac{2}{C|b|}}
%\eeq
%From this one and eqn. (\ref{am}),
%\beq
%r_h
%= r_0 \left[ \cos \left( {\rm am} \left( F(\pi /2,
%1/\sqrt 2 )  - \epsilon \sqrt{\frac{2}{C|b|}} \;\right) \right) \right] ^{-1}
%\eeq
%From the russian table again,
%\beq
%\cos {\rm am} (u) = {\rm cn} (u)
%\eeq
%and then
\beq
r_h = r_0 \left[ {\rm cn} \left( F(\pi /2,
1/\sqrt 2 )  -  \sqrt{\frac{2}{C|b|}} \;\right)\right] ^{-1}
,
\label{rh}
\eeq
where the function cn$(x)$ is the cosine-amplitude. This expression provieds the value of
$r_h$ in terms of $C=2M$ and the constant $|b|$. Note that
$r_h > r_0$ always. Consequently, the region where the geometry is not defined
($r<r_0$) is always inside the surface $r = r_h$.

To determine whether Eqn.(\ref{rh}) defines a horizon or a singular surface, we can compute the
components of the Riemann tensor calculated in
the tetrad system
$$
\omega^t_t = \sqrt{1+\mu (r)},\;\;\;\;\omega^r_r = \sqrt{1+\nu (r)},
$$
$$
\omega^\theta_\theta = r,\;\;\;\;\omega^\phi_\phi = r\sin \theta .
$$
Here we will give the expression for only one of the components, which will
be enough to illustrate the
typical behaviour:
\beq
R_{t\theta t \theta} = \frac{1}{2r} \frac{\mu '(r)}{(1+\nu (r))(1+\mu (r))}
.
\label{riemann}
\eeq
Clearly this component
of the Riemann tensor in the tetrad system depends on the product
$(g_{tt}(r)g_{rr}(r))^{-1}$
which diverges at $r=r_h$ because $g_{tt}(r)$ is null at that point. Consequently, we would
have a naked singular surface instead of a horizon
\footnote{Note that in \Sch geometry, $g_{tt}g_{rr}=-1$, and the singularity at the
surface $r=r_h$ is eased out.}.
We could attempt to cure this
divergence by imposing that $g_{rr}(r)$
diverges at the horizon, in the hope of getting a finite and nonzero value for
the denominator of Eqn.(\ref{riemann}).
Note however that $g_{rr}(r)$ only diverges at $r=r_0$, so it would be necessary
to impose that $r_h=r_0$ to
produce a horizon. To examine the behaviour of the product for $r_h=r_0$,
we expand the metric functions near $r=r_0$. It is convenient to change variables to
$r=r_0 + R$. In terms of this new variable, the significant terms in the
series expansions are
%\beq
$$
g_{tt}(r_0+R) \approx 1 - \frac{C}{8(r_0+R)} \sqrt{\frac{R}{r_0}} + \frac{9C}{8r_0}
\arctan \left(\sqrt{\frac{R}{r_0}} \right),
$$
%\eeq
$$
%\beq
g_{rr}(r_0+R) \approx -1 - \frac{C}{2(r_0+R)}\sqrt{\frac{r_0}{R}} \left( 1 + \frac 5 4
\frac{R}{r_0} \right).
$$
%\eeq
The only term that leads to a divergence in the
limit $R\rightarrow 0$ is the term proportional to $1/\sqrt R$
appearing in the second expression. Taking the product
$g_{tt}(r_0+R)g_{rr}(r_0+R)$ in the limit $R\rightarrow 0$
(and using that $\arctan (x) \approx x - \frac{x^3}{3}$ for small $x$),
it is clear that
the divergence cannot be eliminated. The analysis of the geometry seen by matter
then implies that static and spherically symmetric black holes cannot be described by NDL theory:
the would-be horizon $r=r_h$ is in fact a naked singularity.

\subsection{Effective metric for gravitons}
\label{effmetricg}
We have shown that there are no black holes {\em for matter} in NDL theory.
However, there may be a black hole configuration for gravitons.
As previously mentioned, in NDL theory gravitons interact nonlinearly
with themselves. Consequently, the path of these particles is not governed by
the background metric but by
an effective metric given by \cite{pap1}
$$
\rho_{\mu\nu} = \gamma_{\mu\nu} + \Lambda_{\mu\nu},
$$
with
$$
\Lambda_{\mu\nu} \equiv 2\;\frac{{\cal L}_{AA}}{{\cal L}
_A}\;(F_\mu^{\alpha\beta}F_{\nu(\alpha\beta)}-F_\mu
F_{\nu}),
$$
and ${\cal L}(A-B)$ is an arbitrary Lagrangian.
A short calculation using the Born-Infeld-like
Lagrangian from Eqn.(\ref{bill}) and the metric given by Eqn.(\ref{metric})
shows that
$$
\rho_{00}(r) = 1 - \frac{\nu (r)^2}{b^2r^2+\nu (r)^2},
$$
with $\rho_{11}(r) = -\rho_{00}(r)$, while $\rho_{22}(r)$ and $\rho_{33}(r)$ are as in
Minkowski spacetime. The result for the component of the Riemann
tensor given by Eqn.(\ref{riemann}) calculated with this metric
indicates that there is a (naked) singularity seen only by gravitons at $r=r_0$.
Subsequently, the effective metric for gravitons cannot describe
a \Sch black hole.

\section{Discussion}
\label{disc}
Our analysis revealed that NDL agrees with GR for the three classical tests
performed in the weak field limit. Note however that there are potential
disagreements in situations in which gravity is strong. For instance, for compact objects
like neutron stars, for which typically $M/R\approx 10^5M_\odot/R_\odot$, the convergence of
the series expansions Eqns.(\ref{gtt})-(\ref{d2}) will be much slower, and observable
effects would appear. Another instance in which the two theories differ in their predictions
is in the existence of black holes. As we have shown, there are no solutions in
NDL theory
that describe static
and spherically symmetric black holes. In fact,
from the analysis in Sect.(\ref{ssss}), we conclude that
the two possible cases in the metric seen by matter lead
to naked singularities.

Let us remark that there are two different types of
singularities
appearing in the problem, which we will call geometrical and physical.
For $r_0\neq r_h$,
we encountered two singularities: one located at $r=r_h$ and another
located at $r=r_0$. We can call the first singularity geometrical because
the quantities related to the geometry (like the effective potential and the
Riemann tensor in tetrads) diverge at $r=r_h$,
but the field $F_{\alpha\beta\gamma}$ is finite at that point
(see Eqn.(\ref{f})).
The energy density of the field is also finite:
its energy-momentum tensor is given by
\footnote{Note that there is an additional term in this expression when compared
with Eqn. (31) in Ref.\cite{pap1}.}
%The energy of the field configuration we're dealing with can be calculated
%from the energy-momentum tensor obtained from the standard definition:
%\beq
%T_{\mu\nu}  =
% \frac{2}{\sqrt{-\gamma}}\frac{\delta\sqrt{-\gamma}}{\delta\gamma^{\mu\nu}}
%\eeq
\begin{eqnarray}
T_{\mu\nu} & = &  -{\cal L(A_B)}
\gamma_{\mu\nu} + 2\left[ 2 F_\mu^{\;\alpha\beta} F_{\nu\alpha\beta} +
F_{\alpha\beta\mu}F^{\alpha\beta}_{\;\;\;\nu} - F_\mu F_\nu \right. \nonumber\\
& & \left. - F_{\alpha(\mu\nu)} F^\alpha
\right]- \left [ F^{\alpha (\epsilon\nu)} \varphi_{\alpha\mu} + F^{\alpha (\epsilon\mu)}
\varphi_{\alpha\nu} - F^\alpha _{(\mu\nu )} \varphi_\alpha^\epsilon
\right]_{;\epsilon}. \nonumber
\end{eqnarray}
From this expression, and using the equations of motion, we get for the energy density,
$$
T_{00} = -13\frac{\nu^2}{r^2} +2 \frac{\nu}{r^2}(\mu -\nu 'r) + 2\mu \frac{\nu '}{r},
$$
which is finite at $r=r_h$. On the other hand, at $r=r_0$ the quantities related
to the geometry do not
diverge, but there is a divergence of the field and of the energy density.

The divergences of the geometrical and field quantities are not compatible for $r_0=
r_h$ either.
We expect
that the divergences of the quantities related to the field to be physical. The fact that
the geometry does not display the same pattern of divergences as that
of the field suggests that some
modification should be introduced to the theory.
A possible alternative would be to assume that the (unobservable) background is not
flat spacetime but a curved geometry (for instance, de Sitter spacetime).
Another assumption that must be checked is the postulated coupling
between matter and gravity (given by the metric $g_{\mu\nu} = \gamma_{\mu\nu}+
\varphi_{\mu\nu}$). Work in this direction is currently
under way.

\section*{Acknowledgements}

MN acknowledges CNPq, SEPB acknowledges FAPERJ and
KEH thanks the Ministerio de Educaci\'on y Cultura, Spain, for financial
support.


\begin{thebibliography}{99}

\bibitem{will} See for instance C.Will, Living Rev. Rel. {\bf 4}, 4 (2001).

\bibitem{pap1} M. Novello, V.A. De Lorenci and L.R. de Freitas, Ann. Phys. (NY)
{\bf 254}, 83 (1997).


\bibitem{fey} {\em Lectures on Gravitation}, R. Feynman, Addison-Wesley (1995).

\bibitem{deser} S. Deser, \Grg\/ {\bf 1}, 9 (1970).

\bibitem{vel1}
 M. Novello, V.A. De Lorenci, Luciane R. de Freitas and O.D. Aguiar,
Phys. Lett. A {\bf 254}, 245 (1999), {\tt gr-qc/0002093}.


\bibitem{vel2} {\em Birefringence of Gravitational Waves},
M. Novello, H.J. Mosquera Cuesta and V.A. De Lorenci,
CBPF-NF-039-00, 2000.

\bibitem{vel3} {\em An astrophysical testbed for NDL gravity using gravitational-wave
and neutrino bursts from local supernovae explosions},
H.J. Mosquera Cuesta, M. Novello and V.A. De Lorenci,
CBPF-NF-038-00, 2000.

\bibitem{nlem1} M. Novello, J. Salim, R. Klippert and De Lorenci, Phys. Rev. D {\bf 61}, 045001
(2000), and references therein.

\bibitem{emwh}  F. Baldovin, M. Novello, S.E. Perez Bergliaffa and
J.M. Salim, Class. Quantum Grav. {\bf 17}, 3265 (2000), and references therein.

\bibitem{bi} M. Born and L.Infeld, Nature {\bf 133}, 63 (1934).

\bibitem{rtable} {\em Table of integrals, sums, series and products}, I. Gradshtein
and I. Rizhik (1962).

\bibitem{wald} \emph{General Relativity}, R. M. Wald, U. of Chicago Press (1984).

\bibitem{dinverno} {\em Introducing Einstein's Relativity}, R. D'Inverno, Clarendon Press (1993).

\end{thebibliography}
\end{document}